\documentclass[aps,pre,showpacs,amssymb,math,superscriptaddress,groupaddress]{revtex4}
\usepackage{graphicx}

\usepackage[cp866nav]{inputenc}
\usepackage[english,ukrainian]{babel}

\begin{document}
\def\figurename{Fig.}
\def\tablename{Table}
\title{Mykyta the Fox and networks of language}
\author{Yu. Holovatch}
\email[]{hol@icmp.lviv.ua} \affiliation{Institute for Condensed
Matter Physics, National Academy of Sciences of Ukraine, UA--79011
Lviv, Ukraine} \affiliation{Institut f\"ur Theoretische Physik,
Johannes Kepler Universit\"at Linz, A-4040, Linz, Austria}
\author{V. Palchykov}
\email[]{palchykov@icmp.lviv.ua} \affiliation{Institute for
Condensed Matter Physics, National Academy of Sciences of Ukraine,
UA--79011 Lviv, Ukraine}
\date{9 May 2007}
\begin{abstract}
The results of quantitative analysis of word distribution in two
fables in Ukrainian by Ivan Franko: "Mykyta the Fox" and
"Abu-Kasym's slippers" are reported. Our study  consists of two
parts: the analysis of frequency-rank distributions and the
application of complex networks theory.  The analysis of
frequency-rank distributions shows that the text sizes are enough to
observe statistical properties. The power-law character of these
distributions (Zipf's law) holds in the region of rank variable
$r=20 \div 3000$ with an exponent $\alpha\simeq 1$. This
substantiates the choice of the above texts to analyse typical
properties of the language complex network on their basis. Besides,
an applicability of the Simon model to describe non-asymptotic
properties of word distributions is evaluated.

In describing language as a complex network, usually the words are
associated with nodes, whereas one may give different meanings to
the network links. This results in different network
representations. In the second part of the paper, we give different
representations of the language network and perform comparative
analysis of their characteristics. Our results demonstrate that the
language network of Ukrainian is a strongly correlated scale-free
small world. Empirical data obtained may be useful for theoretical
description of language evolution.

\vspace{1ex} \noindent {\bf Key words:} complex systems, language
networks, scale-free networks, Zipf's law.
\end{abstract}
\pacs{02.10.Ox, 87.75.Da, 89.75.Hc}
 \maketitle

\noindent {\bf \em This is an illustrative material from the paper
submitted in Ukrainian to the Journal of Physical Studies
(http://www.ktf.franko.lviv.ua/JPS/index.html).}

\begin{table}[b*]
\centering
\begin{tabular}{|l|l|l|l|l|l|l|l|}
\hline
$r$&$f$&word&English&$r$&$f$&word&English\\
&& (in Ukrainian)&translation&&&(in Ukrainian)&translation\\
\hline
1&439&я&I&1&165&вўн&he\\
2&323&не&not&2&163&в&in\\
3&312&в&in&3&143&не&not\\
4&272&ў&and&4&140&ў&and\\
5&233&ти&you&5&128&той&those\\
6&222&що&that&6&125&що&that\\
7&214&на&on&7&125&на&on\\
\vdots&\vdots&\vdots&\vdots&\vdots&\vdots&\vdots&\vdots\\
16&140&лис&fox&12&87&капець&slipper\\
21&109&Микита&Mykyta&18&69&Абу-Касим&Abu-Kasym\\
23&98&вовк&wolf&40&28&пан&lord\\
25&88&цар&tsar&41&27&суддя&judge\\
\hline
\end{tabular}
\caption{Rank classification of words from Ivan Franko's "Mykyta the
Fox" \, \cite{Franko1} (left part of the table) and "Abu-Kasym's
slippers"  \,  \cite{Franko2} (right part of the table). $r$: rank,
$f$: number of occurences of a word in the text. The length of the
above texts equals  $ {\cal N} = 15426; \, 8002$, their vocabulary
(number of distinct words) equals  ${\cal V} =3563; \, 2392$ for
"Mykyta the Fox" \, and "Abu-Kasym's slippers" \,  correspondingly.
\label{tab1}}
\end{table}

\begin{figure}[h]
\centerline{\includegraphics[width=80mm]{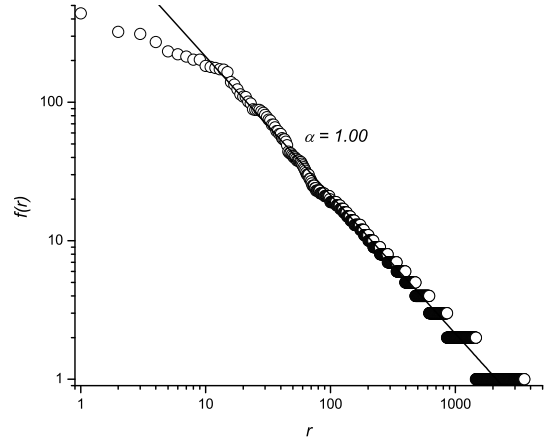}}
\vspace*{3ex} \caption{Frequency-rank dependence for "Mykyta the
Fox". Solid curve: approximation by the power function $f(r)\sim
1/r^\alpha$ with $\alpha=1.00$. Similar dependencies result for
"Abu-Kasym's slippers" \, and for both texts joined together (with
the values of $\alpha=0.97$ and $\alpha=1.00$, correspondingly).
Typical accuracy of $\alpha$ is $\chi ^2/d.o.f=0.002$.\label{fig1}}
\end{figure}

\begin{figure}[h]
\centerline{\includegraphics[width=95mm,angle=0]{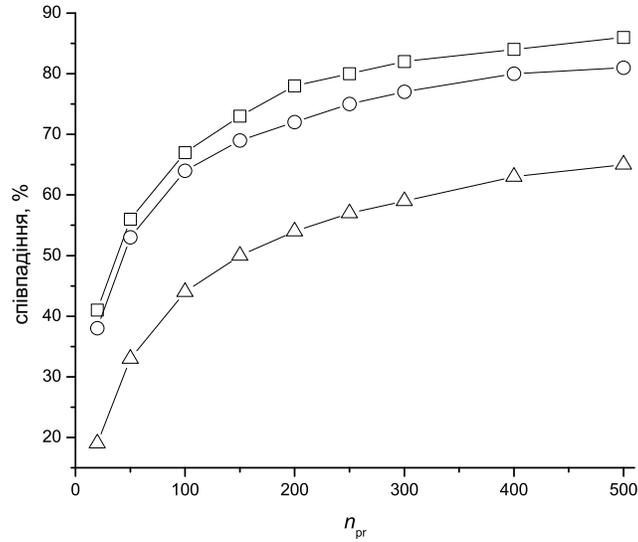}}
\vspace*{3ex} \caption{Results of comparison of computer-generated
and original texts. The curves show dependencies of the percent of
coincidence of two texts as a function of predicted word block
$n_{\rm pr}$. $\circ$: comparison of the original text with the text
generated according to the Simon model. $\triangle$: comparison the
text generated according to the Simon model with the randomly
generated text. $\Box$: comparison of two texts generated according
to the Simon model. \label{fig2}}
\end{figure}

\begin{figure}[h]
\centerline{\includegraphics[width=130mm]{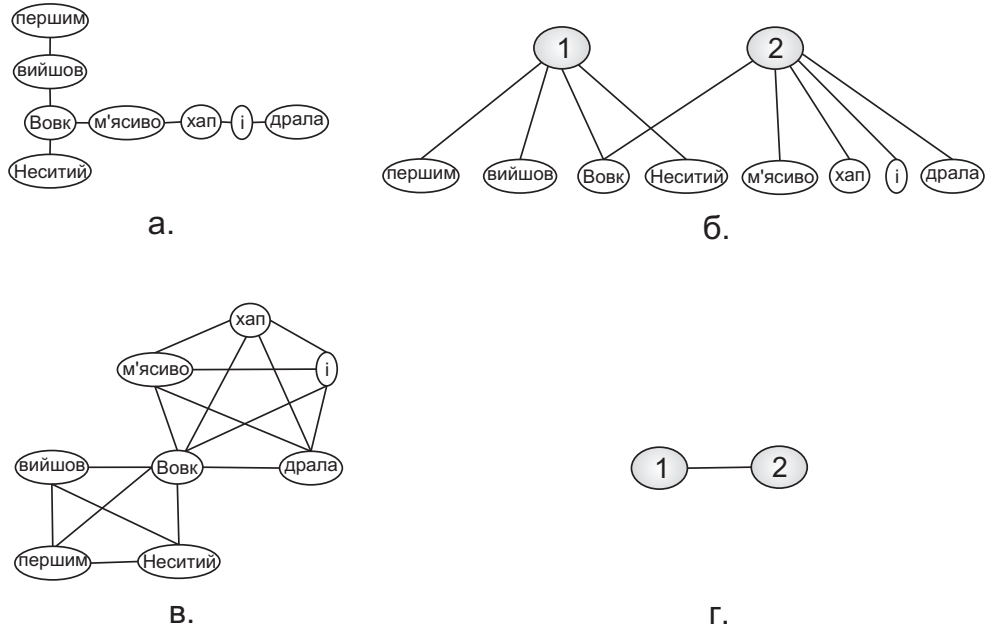}} \vspace*{3ex}
 \caption{Representation of two sentences, 1: "Першим
вийшов Вовк Неситий", 2: "Вовк м'ясиво хап -- ў драла!"\, in a form
of graphs. {\bf a.} $\mathbb{L}$-space. Links connect neighbouring
words, that belong to the same sentence. A number of neighbours for
each word (word window) is defined by the "radious of interaction"
\, $1 \leq R \leq R_{\rm max}$. In the given example $R=1$. For
$R=1$ only the neighbouring words in a sentence are connected, for
$R=2$ links connect nearest and next nearest neighbours in a
sentence, $R=R_{\rm max}$ corresponds to the sentence length.  {\bf
б.} $\mathbb{B}$-space. Nodes of two sorts are present. Dark nodes:
sentences, light nodes: words that belong to them. {\bf в.}
$\mathbb{P}$-space. All words, that belong to the same sentence are
connected. {\bf г.} $\mathbb{C}$-space. Sentences are connected if
they contain the same words. A link between nodes-sentences (1 and
2) corresponds to the word "вовк", common for both sentences. For
the nomenclature of the above spaces see Ref. \cite{vonFerber06}.
For the language networks, the $\mathbb{L}$-space representation was
introduced in Ref. \cite{Ferrer01} and the $\mathbb{P}$-space
representation was introduced in Ref. \cite{Caldeira05}. Note that
at $R=R_{\rm max}$ $\mathbb{L}$-space representation coincides with
the $\mathbb{P}$-space representation. \label{fig3}}
\end{figure}

 \begin{table}[th]
\centering

\begin{tabular}{|c|r|r|c|r|c|c|c|c|r|c|c|}
\hline
 $R$&${\cal V}$&${\cal M}$&$\langle{k}\rangle$&$\langle{k^2}\rangle/\langle{k}\rangle$&
 $k_{\rm max}$&$\gamma$&$\gamma_{\rm int}$&$\langle{C}\rangle$&$\langle{C}\rangle/
 {C_r}$&$\langle{l}\rangle$&$l_{\rm max}$\\
\hline
%Абу-Касимовў капцў
1&2392&6273&5.24&48&228&1.9&1.15&0.171866&78&3.428&11\\
 2&2392&11475&9.59&77&391&2.0&1.18&0.567195&141&2.897&7\\
 $R_{\rm max}$&2392&48603&40.64&208&1134&1.9&1.35&0.841215&50&2.220&4\\
 \hline
 %Лис Микита
 1&3563&11102&6.23&76&419&1.9&1.12&0.214024&122&3.301&11\\
 2&3563&20063&11.26&119&665&1.8&1.16&0.587752&186&2.852&7\\
 $R_{\rm max}$&3563&65997&37.05&269&1526&1.9&1.27&0.821895&79&2.274&5\\
 \hline
 %Загально
 1&4823&16580&6.88&102&537&1.9&1.13&0.243097&170&3.235&11\\
 2&4823&29916&12.41&156&868&1.8&1.15&0.585375&227&2.826&7\\
 $R_{\rm
 max}$&4823&107750&44.68&360&2185&2.0&1.28&0.818495&88&2.249&5\\
 \hline
\end{tabular}
\caption{Quantitative characteristics of word networks for the texts
under consideration in $\mathbb{L}$-space for several values of $R$.
Upper part of the table: word network for the text "Abu-Kasym's
slippers" \cite{Franko2}, middle part: "Mykyta the Fox"
\cite{Franko1}, lower part: both texts together. ${\cal V}$: number
of nodes, ${\cal M}$: number of links, $\langle{k}\rangle$, $k_{\rm
max}$: mean and maximal node degree, $\gamma$, $\gamma_{\rm int}$:
exponents of the node degree ($ P(k)\sim k^{-\gamma}$) and of the
cumulative node degree ($P_{\rm int}(k)=\sum_{k^{\prime}=k}^{k_{\rm
max}}P(k^{\prime})$) distributions, $\langle{C}\rangle$: mean value
of the clustering coefficient, $C_r$: clustering coefficient of the
classical Erd\"os-R\'enyi random graph of the same size,
$\langle{l}\rangle$, $l_{\rm max}$: mean and maximal values of the
shortest path length. Note, that at $R=R_{\rm max}$ representation
of a network in $\mathbb{L}$- and $\mathbb{P}$-spaces do coincide.
\label{tab2}}
\end{table}

\begin{figure}[h]
\centerline{\includegraphics[width=80mm]{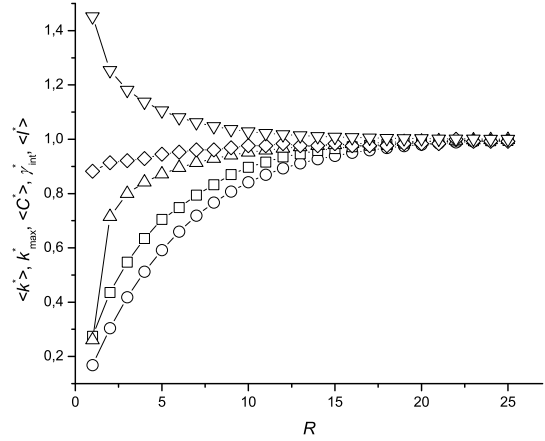}}
\caption{"Mykyta the Fox": dependence of word network
characteristics on the size of a word window $R$. Mean ($\langle
k^*\rangle$, $\circ$-$\circ$-$\circ$) and maximal ($k^*_{\rm max}$,
$\Box$-$\Box$-$\Box$) node degrees, mean clustering coefficient
($\langle C^*\rangle$, $\triangle$-$\triangle$-$\triangle$) and
shortest path length ($\langle \l^*\rangle$,
$\nabla$-$\nabla$-$\nabla$), cumulative node degree distribution
exponent ($\gamma^*_{\rm int}$,
$\diamondsuit$-$\diamondsuit$-$\diamondsuit$) are normalized by
their values at $R=R_{\rm max}$. An increase of  $R$ causes an
increase of number of links in the network. This is the reason for
an increase  of $\langle C\rangle$, $\langle k\rangle$, $\langle
k_{max}\rangle$ and for a decrease of $\langle l\rangle$ with $R$.
\label{fig4}}
\end{figure}

\begin{figure}[h]
\centerline{\includegraphics[width=80mm]{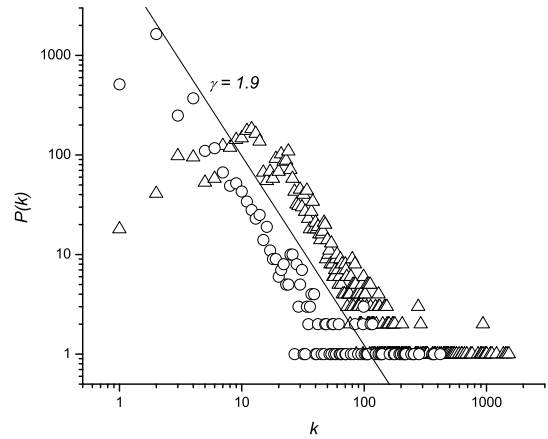}}
\caption{Node degree distribution for "Mykyta the Fox" \, follows a
power law $P(k)\sim 1/k^\gamma$ for different $R$ ($R = 1$
($\circ$-$\circ$-$\circ$), $R = R_{\rm max}$ (
$\triangle$-$\triangle$-$\triangle$)). A difference between the
exponents of the power law for different $R$ can not be
distinguished for the texts under consideration. Solid line shows a
power law with an exponent $\gamma=1.9$.\label{fig5}}
\end{figure}

\begin{figure}[h]
\centerline{\includegraphics[width=80mm]{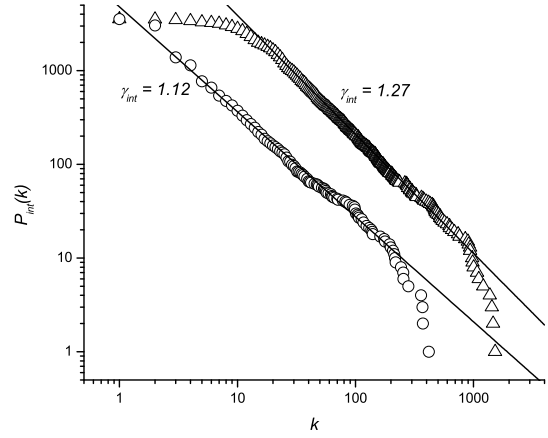}}
\vspace*{3ex} \caption{Cumulative node degree distribution for
"Mykyta the Fox" \, also shows a power law dependence. Within an
accuracy of the plot one can see an increase of the  exponent
$\gamma_{\rm int}$ with $R$. $\gamma_{\rm int}=1.12$ for $R = 1$
($\circ$-$\circ$-$\circ$), $\gamma=1.27$  for $R = R_{\rm max}$
($\triangle$-$\triangle$-$\triangle$). \label{fig6}}
\end{figure}

\begin{figure}[h]
\centerline{\includegraphics[width=80mm]{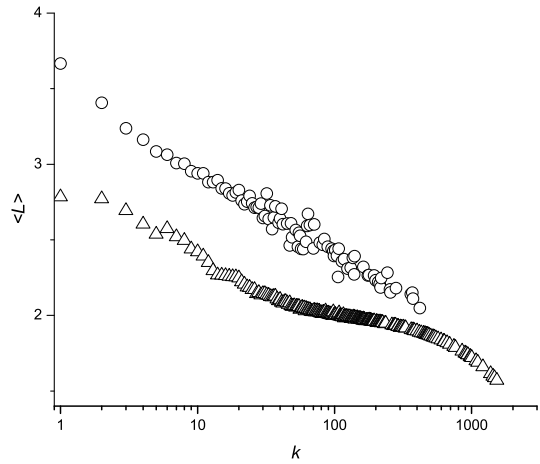}}
\vspace*{3ex} \caption{"Mykyta the Fox": mean shortest path length
from the node of degree $k$ to the rest of network nodes. $R = 1$
($\circ$-$\circ$-$\circ$), $R = R_{\rm max}$
($\triangle$-$\triangle$-$\triangle$). Decrease of $\langle
l\rangle$ with $k$ indicates that hubs (most connected nodes) are in
a closer reach from each other than other nodes. Very small value of
$\langle l\rangle$ indicates that  this network is a small world.
\label{fig7}}
\end{figure}

\begin{figure}[h]
\centerline{\includegraphics[width=80mm]{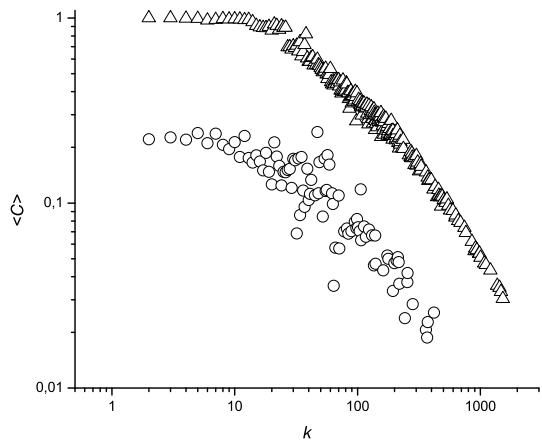}}
\vspace*{3ex} \caption{Mean clustering coefficient as a function of
the node degree for "Mykyta the Fox". $R = 1$
($\circ$-$\circ$-$\circ$), $R = R_{\rm max}$
($\triangle$-$\triangle$-$\triangle$). These dependencies are
characterized by a plateau at small values of $k$ and further
decreasing. An increase of the mean clustering coefficient  with $R$
is explained by an increase of the number of links with $R$ at
constant number of nodes. For the similar reason, $\langle C\rangle$
increases with the text length. \label{fig8}}
\end{figure}


\clearpage
\begin{thebibliography}{50}

\bibitem{Franko1} Ivan Franko, {\em Mykyta the Fox}, (Works in 50 volumes, vol. 4,
p. 57, Naukova Dumka, Kyiv, 1976), in Ukrainian. Available on-line
at: http://poetyka.uazone.net/franko/

\bibitem{Franko2}
Ivan Franko, {\em Abu-Kasym Slippers}, (Works in 50 volumes, vol. 4,
p. 295, Naukova Dumka, Kyiv, 1976), in Ukrainian. Available on-line
at: http://poetyka.uazone.net/franko/

\bibitem{exotic_books}
 G.~Parisi, {\em Complex Systems: a Physicist's Viewpoint}
 (preprint cond-mat/0205297, 2002);
%\bibitem{Mantegna99}
 R.~N.~Mantegna, H.~E.~Stanley, {\em An Introduction to
Econophysics: Correlations and Complexity in Finance} (Cambridge
University Press, Cambridge, 1999);
%\bibitem{Chakrabarti06}
 B.~K.~Chakrabarti, A.~Chakraborti, A.~Chatterjee, {\em Econophysics and
Sociophysics: Trends and Perspectives} (Wiley-VCH, Berlin, 2006).

\bibitem{Dorogovtsev03} S.~N.~Dorogovtsev, S.~N.~Mendes, {\em Evolution
of Networks} (Oxford University Press, Oxford, 2003).

\bibitem{Holovatch06}
Yu.~Holovatch, A.~Olemskoi, C.~von~Ferber, O.~Mryglod, T.~Holovatch,
I.~Olemskoi, V.~Palchykov, J. Phys. Stud. {\bf 10}, (2006).

\bibitem{critical_books}
H.~E.~Stanley, {\em Introduction to Phase Transitions and Critical
Phenomena} (Clarendon Press, Oxford, 1971);
%\bibitem{Domb96}
C.~Domb. {\em The Critical Point} (Taylor \& Francis, London
Bristol, 1996).

\bibitem{Holovatch04}
Yu.~Holovatch (Ed.), {\em Order, Disorder and Criticality. Advanced
Problems of Phase Transition Theory} (World Scientific, Singapore,
2004).

\bibitem{graphs}
{\em Graph Theory}, (Springer-Verlag, Heidelberg, Graduate Texts in
Mathematics, Volume 173, 2005); S.~Bornholdt, H.~Schuster (Eds.),
{\em Handbook of Graphs and Networks} (Wiley-VCH, Weinheim, 2003).

\bibitem{network_reviews}
%\bibitem{Albert02}
R.~Albert, A.-L.~Barab\'asi, Rev. Mod. Phys. {\bf 74}, 47 (2002);
%\bibitem{Dorogovtsev02}
S.~N.~Dorogovtsev, J.~F.~F.~Mendes, Adv. Phys. {\bf 51}, 1079
(2002);
%\bibitem{Newman03}
M.~E.~J.~Newman, SIAM Review {\bf 45}, 167 (2003);
%\bibitem{Boccaletti06}
S.~Boccaletti, V.~Latora, Y.~Moreno, M.~Chavez, D.-U.~Hwang, Physics
Reports {\bf 424}, 175 (2006);
%\bibitem{Lesne06}
A.~Lesne, Lett. Math. Phys. {\bf 78}, 235 (2006).

\bibitem{network_books}
%\bibitem{Watts99}
D.~J.~Watts, {\em Small Worlds} (Princeton University Press,
Princeton, NJ, 1999);
%\bibitem{Pastor04}
R.~Pastor-Satorras, A.~Vespignani, {\em Evolution and Structure of
the Internet: A Statistical Physics Approach} (Cambridge University
Press, Cambridge, 2004);
%\bibitem{Newman06}
M.~E.~J.~Newman, A.-L.~Barab\'asi,  D.~J.~Watts,
{\em The Structure and Dynamics of Networks} (Princeton University
Press, Princeton, 2006).

\bibitem{Watts98} D.~J.~Watts, S.~H.~Strogatz, Nature (London) {\bf 393}, 440
(1998).

\bibitem{Albert99}
R.~Albert, H.~Jeong,  A.-L.~Barab\'asi, Nature (London) {\bf 401},
130 (1999).

\bibitem{Ferrer01} R.~Ferrer~i~Cancho, R.~V.~Sol\`e, Proc. R. Soc.
Lond. B {\bf 268}, 2261 (2001).

\bibitem{Dorogovtsev01} S.~N.~Dorogovtsev, J.~F.~F.~Mendes, Proc. R. Soc.
Lond. B {\bf 268}, 2603 (2001).

\bibitem{Caldeira05} S.~M.~G.~Caldeira, T.~C.~Petit Lobao, R.~F.~S.~Andrade,
A.~Neme, J.~G.~V.~Miranda, preprint physics/0508066 (2005).

\bibitem{Ferrer04a} R.~Ferrer~i~Cancho, R.~V.~Sol\'e, R.~Kohler, Phys.
Rev. E {\bf 69}, 051915 (2004).

\bibitem{Ferrer04b} R.~Ferrer~i~Cancho, Phys. Rev. E {\bf 70}, 056135 (2005).

\bibitem{Ferrer05a} R.~Ferrer~i~Cancho, O.~Riordan, B.~Bollob\'as, Proc. R. Soc.
Lond. B {\bf 272}, 561 (2005).

\bibitem{Sole05} R.~Sol\'e, Nature {\bf 434}, 289 (2005).

\bibitem{Motter02} A.~E.~Motter, A.~P.~S.~de~Moura, Y.-C.~Lai, P.~Dasgupta,
Phys. Rev. E {\bf 65}, 065102(R) (2002).

\bibitem{Sigman02} M.~Sigman, G.~A.~Cecchi, Proc. Natl. Acad. Sci. USA, {\bf
99}, 1742 (2002).

\bibitem{Jesus04} A.~de~Jesus Holanda, I.~Torres Pisa, O.~Kinouchi,
A.~Souto Martinez, E.~E.~Seron Ruiz, Physica A {\bf 344}, 530
(2004).

\bibitem{Zipf49} G.~F.~Zipf, {\em Human Behaviour and the Principle of least Effort.
An Introduction to Human Ecology}, 1st edition (Hafner reprint, New
York, 1972) (Addison-Wesley, Cambridge, MA, 1949).

\bibitem{Zipf35} G.~F.~Zipf, {\em The Psycho-Biology of Language}, (Houghton-Mifflin,
Boston, 1935).

\bibitem{Zipfsite} Bibliography about Zipf's law may be found at: http://www.nslij-genetics.org/wli/zipf/

\bibitem{Ferrer05b} R.~Ferrer~i~Cancho, Eur. Phys. J. B {\bf 44}, 249
(2005).

\bibitem{Mitzenmacher02}
M.~Mitzenmacher, Internet Mathematics {\bf 1}, 226 (2004).

\bibitem{Simkin06}
M.~V.~Simkin, V.~P.~Roychowdhury, preprint physics/0601192 (2006).

\bibitem{Condon28}
E.~U.~Condon, Science {\bf 67}, 300 (1928). An author of this paper
about statistics of the word distribution in a dictionary is Edward
Uhler Condon (1902-1974), an author of the Franck-Condon principle
and of the first paper about quantum mechanics in English (together
with Philip Morse in 1929).

\bibitem{Pavlov01}
A.~N.~Pavlov, W.~Ebeling, L.~Molgedey, A.~R.~Ziganshin,
V.~S.~Anishchenko, Physica A {\bf 300}, 310 (2001).

\bibitem{Montemuro01} M.~A.~Montemuro, Physica A {\bf 300}, 567
(2001).

\bibitem{Dahui05} W.~Dahui, L.~Menghui, D.~Zengru, Physica A {\bf
358}, 545 (2005).

\bibitem{Montemuro02} M.~A.~Montemuro, D.~H.~Zanette, Advances in Complex Systems  {\bf 5}, 7
(2002).

\bibitem{Kokol00}
P.~Kokol, V.~Podgorelec, Complexity International {\bf 7}, 1 (2000).

\bibitem{Kanter95}
I.~Kanter, D.~A.~Kessler, Phys. Rev. Lett. {\bf 74}, 4559 (1995).

\bibitem{Melnyk05}
S.~S.~Melnyk, O.~V.~Usatenko, V.~A.~Yampol'skii, V.~A.~Golick, Phys.
Rev. E {\bf 72}, 026140 (2005).

\bibitem{Simon55}
H.~A.~Simon, Biometrika {\bf 42}, 425 (1955).

\bibitem{Abramowitz79} M. Abramowitz, I.A. Stegun (Eds.),
{\em Handbook of Mathematical Functions} (National Bureau of
Standards, 1964).

\bibitem{transport}
V.~Latora, M.~Marchiori, Physica A {\bf 314}, 109 (2002); P.~Sen,
S.~Dasgupta, A.~Chatterjee, P.~A.~Sreeram, G.~Mukherjee,
S.~S.~Manna, Phys. Rev. E {\bf 67}, 036106 (2003); K.~A.~Seaton,
L.~M.~Hackett, Physica A {\bf 339}, 635 (2004); J.~Sienkiewicz,
J.~A.~Holyst, Phys. Rev. E {\bf 72}, 046127 (2005).

\bibitem{vonFerber06} C. von Ferber, T. Holovatch, Yu. Holovatch, V. Palchykov,
Physica A (2007), doi:10.1016/j.physa.2007.02.101; preprint
physics/0608125 (2006).

\bibitem{bnc} The British National Corpus is a 100 million word collection of
samples of written and spoken language from a wide range of sources,
designed to represent a wide cross-section of current British
English, both spoken and written. See: http://www.natcorp.ox.ac.uk/

\bibitem{Milgram67}
S.~Milgram, Psychol. Today {\bf 2}, 60 (1967).

\bibitem{Gutenberg} Analyzed in Ref. \cite{Caldeira05} texts mainly are taken from
the site of Gutenberg project: http://www.gutenberg.org/

\bibitem{Nowak99} M.~A.~Nowak, D.~C.~Krakauer, Proc. Natl. Acad. Sci. USA, {\bf
96}, 8028 (1999).

\end{thebibliography}
\end{document}